\documentclass{ws-p8-50x6-00}

\begin{document}
\title{Selectron Pair Production in $e^-e^-$ Scattering}

\author{C. Bl\"ochinger, H. Fraas and T. Mayer}

\address{University of W\"urzburg, Am Hubland, 97074 W\"urzburg, Germany\\ 
E-mail: bloechi@physik.uni-wuerzburg.de, fraas@physik.uni-wuerzburg.de, mayer@physik.uni-wuerzburg.de}




\maketitle

\abstracts{
We investigate pair production of selectrons in $e^-e^-$ scattering with subsequent decay 
into an electron and the LSP including ISR and beam\-strahlung. This process can be used at a 
linear collider to measure the selectron
masses and the gaugino mass parameter $M_1$ very precisely.}

\section{Introduction}
The determination of the masses and couplings of supersymmetric particles will be one of the main
aims of linear colliders in the 500 GeV - 3 TeV range \cite{TESLA,NLC,JLC,CLIC}. 
For a precission measurement of the selectron masses the $e^-e^-$ mode is of particular interest 
\cite{Peskin1,Peskin2,Zerwas1}. The
main advantage is the $\beta$ dependence of the total cross sections near threshold for the production
of two selectrons associated to the same helicity in contrast to the $\beta^3$ behavior in $e^+e^-$
annihilation. A further advantage is the higher polarization of $e^-$ beams compared to $e^+$ beams.

The production process $e^-e^-\rightarrow\tilde{e}^-_{L/R}\tilde{e}^-_{L/R}$ proceeds via exchange
of all four neutralinos in the t-channel. Since at least for $M_1<1$ TeV one neutralino mass depends 
strongly on $M_1$ this process can be used to determine the 
gaugino mass parameter $M_1$ \cite{bloechi1}.

\begin{figure}[htb]
\label{fig:thresh}
\centering
\begin{picture}(11.8,4.3)
\put(0.0,-0.9){\includegraphics{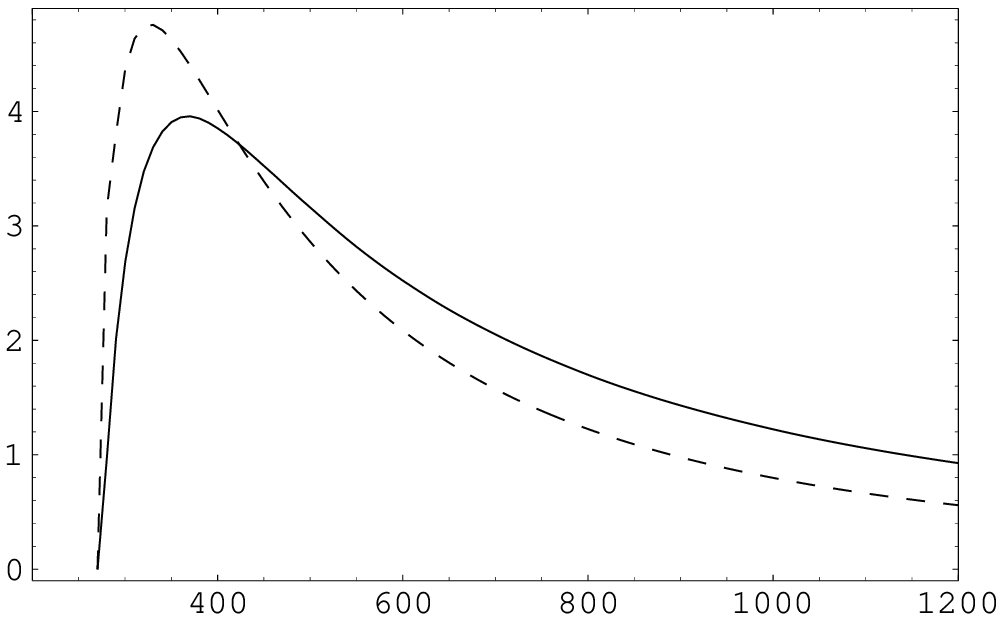}}
\put(4.8,0.05){{\tiny $\sqrt{s}$/GeV}}
\put(0.0,4.){{\tiny $\sigma_{ee}$/pb}}
\put(5.9,-0.9){\includegraphics{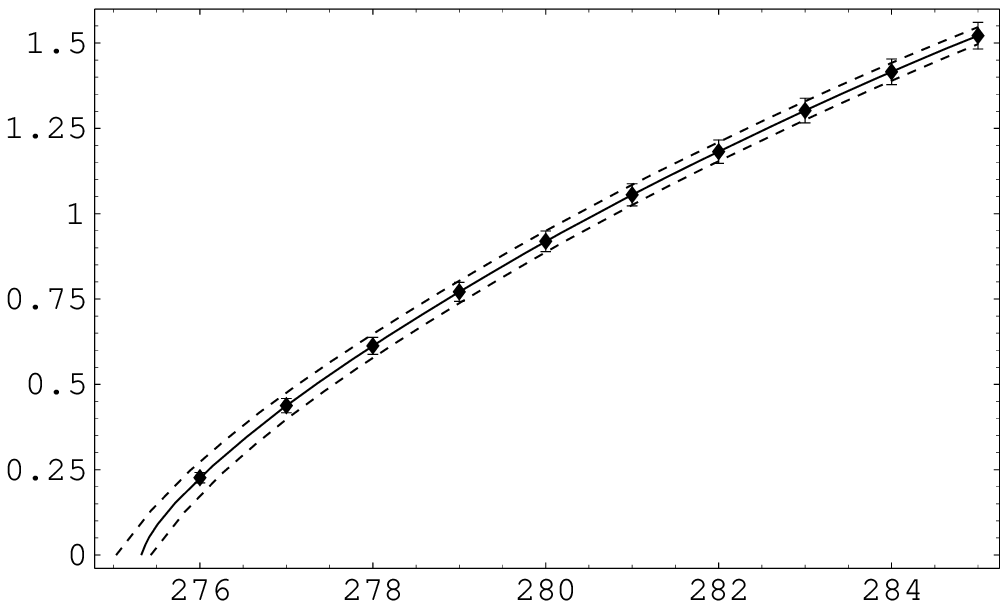}}
\put(10.9,0.05){{\tiny $\sqrt{s}$/GeV}}
\put(5.9,4.0){{\tiny $\sigma_{ee}$/pb}}
\put(2.6,4.0){{\tiny (a)}}
\put(8.9,4.0){{\tiny (b)}}
\end{picture}
\caption{(a) Total cross section $\sigma_{ee}$ of the process 
$e^-e^-\rightarrow\tilde{e}^-_{L/R}\tilde{e}^-_{L/R}\rightarrow\tilde{\chi}_1^0\tilde{\chi}_1^0e^-e^-$
with longitudinal polarizations $P_{e1}=P_{e2}=80\%$, supersymmetric parameters $M_2=152$ GeV, $\mu=316$ GeV, $\tan\beta=3$ and selectron masses $m_{\tilde{e}_R}=137.7$ GeV, $m_{\tilde{e}_L}=179.3$ GeV without (- - -) and with (----) ISR and beamstrahlung. (b) Same as (a) at the $\tilde{e}^-_{R}\tilde{e}^-_{R}$ threshold with ISR and beamstrahlung. The error bars are calculated assuming $\mathcal{L}=1$ fb$^{-1}$ for each point. The dotted lines are the total cross sections for $m_{\tilde{e}_R}\pm 100$ MeV.}
\end{figure}

\section{Mass Determination at Threshold}

In fig.~\ref{fig:thresh}a we show the total cross section $\sigma_{ee}$ (summed over all kinematical reachable selectron combinations) of the process
$e^-e^-\rightarrow\tilde{e}^-_{L/R}\tilde{e}^-_{L/R}\rightarrow\tilde{\chi}_1^0\tilde{\chi}_1^0e^-e^-$
for longitudinal beam polarizations $P_{e_1}=P_{e_2}=80\%$, supersymmetric parameters 
$M_2=152$ GeV, $\mu=316$ GeV and $\tan\beta=3$ and  selectron masses 
$m_{\tilde{e}_R}=137.7$ GeV, $m_{\tilde{e}_L}=179.3$ GeV. The steep rise with $\beta$ of 
$\sigma_{ee}$ at the $\tilde{e}_R^-\tilde{e}_R^-$ threshold is not substantially flattened
 if we include 
ISR \cite{Jadach1} and beamstrahlung
\cite{beam1}. For this choice of polarization $\tilde{e}_R^-\tilde{e}_L^-$ and $\tilde{e}_L^-\tilde{e}_L^-$
production is suppressed.
For the same parameters and beam polarizations the cross section at the threshold for the production of two 
right selectrons including ISR and beamstrahlung corrections 
is shown in  fig.~\ref{fig:thresh}b. For the chosen parameter point the right
selectron decays completely into an electron and the LSP.
If we assume, that the cross section will be measured at ten points between $\sqrt{s_{ee}}=276$ GeV
and $\sqrt{s_{ee}}=286$ GeV,  c.f. fig.~\ref{fig:thresh}b, with a luminosity $\mathcal{L}=1$~fb$^{-1}$
at each point, the right selectron mass can be determined with
 a statistical error $<(\pm 100\mbox{ MeV})$. This can be seen
in fig.~\ref{fig:thresh}b, where the total cross section for selectron masses $m_{\tilde{e}_R}\pm 100$ MeV
is shown as dotted lines. 

\section{$M_1$ Determination}

\begin{figure}[htb]
\label{fig:M1}
\centering
\begin{picture}(11.8,4.3)
\put(0.0,-0.9){\includegraphics{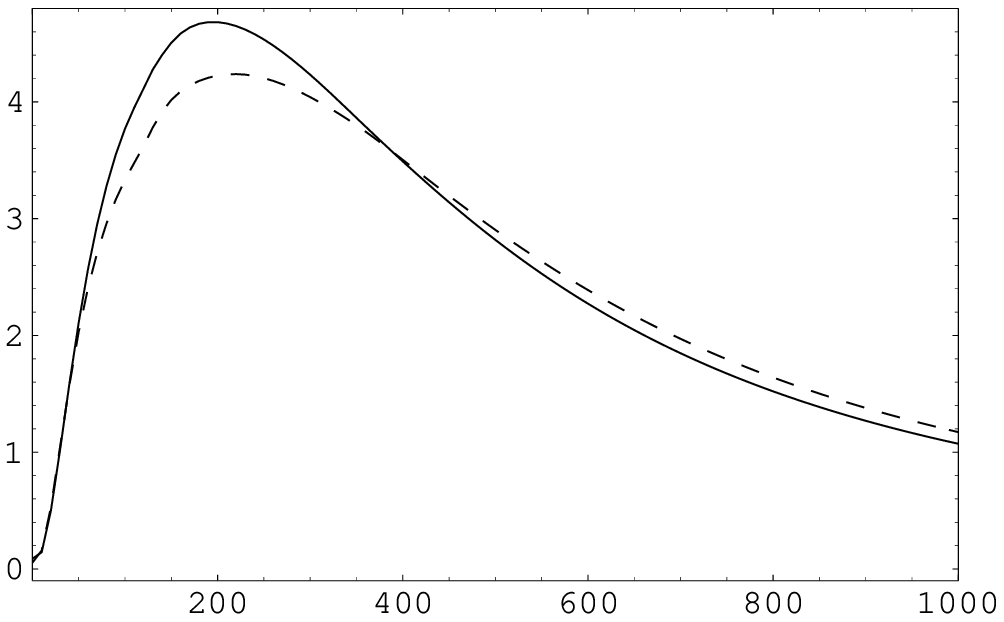}}
\put(4.8,0.05){{\tiny $M_1$/GeV}}
\put(0.0,4.){{\tiny $\sigma_{ee}$/pb}}
\put(5.9,-0.9){\includegraphics{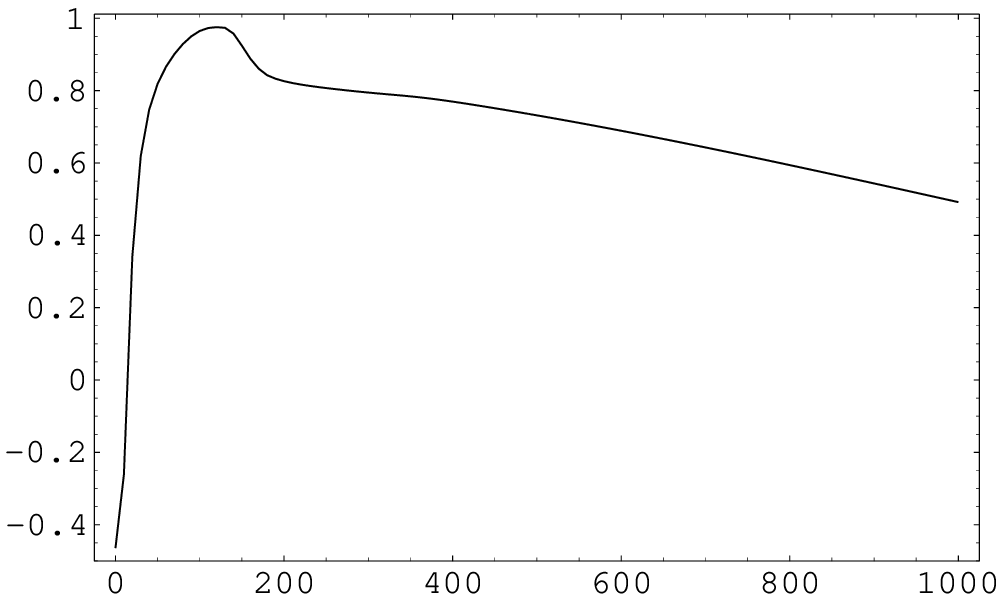}}
\put(10.9,0.05){{\tiny $M_1$/GeV}}
\put(5.9,4.0){{\tiny $A_{pol}$}}
\put(2.6,4.0){{\tiny (a)}}
\put(8.9,4.0){{\tiny (b)}}
\end{picture}
\caption{(a) Dependence of the total cross section $\sigma_{ee}$ of the process 
$e^-e^-\rightarrow\tilde{e}^-_{L/R}\tilde{e}^-_{L/R}\rightarrow\tilde{\chi}_1^0\tilde{\chi}_1^0e^-e^-$
with longitudinal polarizations $P_{e_1}=P_{e_2}=80\%$, supersymmetric parameters $M_2=152$ GeV, $\mu=316$ GeV, $\tan\beta=3$ and selectron masses $m_{\tilde{e}_R}=137.7$ GeV, $m_{\tilde{e}_L}=179.3$ GeV at $\sqrt{s}=500$ GeV without (- - -) and with (----) ISR and beamstrahlung. (b) Polarization asymmetry $A_{pol}$ with ISR and beamstrahlung as defined in (\ref{apol}) for the same parameters as in (a).}
\end{figure}

If we assume unification of the gaugino masses at the GUT-scale the relation
$M_1=5/3\tan^2\theta_W\times M_2$ can be derived by solving renormalisation group equations. 
This relation can be tested with selectron pair production and decay, because 
the masses and couplings of the four exchanged neutralinos strongly depend on $M_1$ 
\cite{bloechi1,bloechi2}.
In fig.~\ref{fig:M1}a we show the $M_1$-dependence of the total cross section $\sigma_{ee}$ of the process
$e^-e^-\rightarrow\tilde{e}^-_{L/R}\tilde{e}^-_{L/R}\rightarrow\tilde{\chi}_1^0\tilde{\chi}_1^0e^-e^-$
for longitudinally polarized beams $P_{e_1}=P_{e_2}=80\%$ and supersymmetric parameters  
$M_2=152$ GeV, $\mu=316$ GeV, $\tan\beta=3$, $m_{\tilde{e}_R}=137.7$ GeV and  $m_{\tilde{e}_L}=179.3$ GeV 
at $\sqrt{s}=500$ GeV. The high values of the cross section and the strong dependence on $M_1$ for
$M_1<1$ TeV can be used to determine $M_1$ very precisely with an ambiguity, however. Note that the strong
$M_1$ dependence of $\sigma_{ee}$ is not substantially modified by ISR and beamstrahlung. The ambiguity
can be resolved by measuring the polarization
asymmetry 
\begin{equation}\label{apol}
A_{pol}= \frac{\sigma_{ee}\left(P_{e_{1/2}}=80\%\right)-\sigma_{ee}\left(P_{e_{1/2}}=-80\%\right)}{\sigma_{ee}\left(P_{e_{1/2}}=80\%\right)+\sigma_{ee}\left(P_{e_{1/2}}=-80\%\right)}\nonumber
\end{equation}
shown in fig.~\ref{fig:M1}b, which again includes ISR and beamstrahlung and also shows a strong 
dependence on $M_1$. Since both observables show a strong and clear $M_1$-dependence
in a large region the $e^-e^-$ mode is more appropriate for the determination of $M_1$ than other
linear collider modes \cite{bloechi1}.

\section*{Acknowledgments}
This work was supported by the Deutsche Forschungsgemeinschaft, contract FR 1064/4-1 and the 
Bundesministerium f\"ur Bildung und Forschung, contract 05 HT9WWa 9.

\end{document}